   \documentstyle[12pt]{article}
\if@twoside\oddsidemargin25mm
\else\oddsidemargin25mm\evensidemargin25mm\marginparwidth25mm\fi%
\begin{document}
\newcommand{\ft}[2]{{\textstyle\frac{#1}{#2}}}
\newcommand{\QED}{{\hspace*{\fill}\rule{2mm}{2mm}\linebreak}}
\newcommand{\be}{\begin{equation}}
\newcommand{\ee}{\end{equation}}
\newcommand{\ba}{\begin{eqnarray}}
\newcommand{\ea}{\end{eqnarray}}
\newtheorem{definizione}{Definition}[section]
\newcommand{\bd}{\begin{definizione}}
\newcommand{\ed}{\end{definizione}}
\newtheorem{teorema}{Theorem}[section]
\newcommand{\bth}{\begin{teorema}}
\newcommand{\eth}{\end{teorema}}
\newtheorem{lemma}{Lemma}[section]
\newcommand{\blem}{\begin{lemma}}
\newcommand{\elem}{\end{lemma}}
\newcommand{\brr}{\begin{array}}
\newcommand{\err}{\end{array}}
\newcommand{\nn}{\nonumber}
\newtheorem{corollario}{Corollary}[section]
\newcommand{\bcorol}{\begin{corollario}}
\newcommand{\ecorol}{\end{corollario}}
\def\twomat#1#2#3#4{\left(\begin{array}{cc}
 {#1}&{#2}\\ {#3}&{#4}\\
\end{array}
\right)}
\def\twovec#1#2{\left(\begin{array}{c}
{#1}\\ {#2}\\
\end{array}
\right)}
\def\Z{{\bf Z}}
\def\R{{\bf R}}
\def\C{{\bf C}}
\def\I{{\bf I}}
\def\h{{\bf h}}
\def\k{{\bf k}}
\def\g{{\bf g}}
\begin{titlepage}

\hfill
\vbox{\hbox{CERN-TH/99-53}\hbox{hep-th/990000}\hbox{March, 1999}}
\vfill
\vskip 3cm
\begin{center}
{\LARGE {  On Central Charges and Hamiltonians\\
for 0-brane dynamics}$^*$}\\
\vskip 1.5cm
  {\bf
R. D'Auria$^1$ S. Ferrara$^2$ and M.A. Lled\'o$^1$
 } \\
\vskip 0.5cm
{\small
$^1$ Dipartimento di Fisica, Politecnico di Torino,\\
 Corso Duca degli Abruzzi 24, I-10129 Torino.\\
and Istituto Nazionale di Fisica Nucleare (INFN) \\ Sezione di Torino, Italy.\\
\vspace{6pt}
$^2$ CERN Theoretical Division, CH 1211 Geneva 23, Switzerland.}
\end{center}
\vskip 4cm


  {\small

\begin{abstract}
We consider general properties of central charges of zero-branes and associated duality
 invariants, in view of their double role, on the bulk and on the world volume (quantum
 mechanical) theory.

A detailed study of the BPS condition for the mass spectrum arising from toroidal compactifications
is given for 1/2, 1/4 and 1/8
 BPS states in any dimension. As a byproduct, we retrieve the U-duality invariant conditions
 on the charge (zero mode) spectrum and the orbit classification of BPS states preserving
 different fractions of supersymmetry.

The BPS condition for 0-branes in theories with 16 supersymmetries in
any dimension is also  discussed.
\end{abstract}  }
  \vspace{2mm} \vfill \hrule width 3.cm
{\footnotesize
 $^*$ Supported in part by   EEC  under TMR contract
 ERBFMRX-CT96-0045,(LNF Frascati,
Politecnico di Torino) and by DOE grant
DE-FGO3-91ER40662}
 \end{titlepage}
\eject
\section{Introduction.}

In recent time, the role of duality symmetries of a dynamical theory
encompassing quantum gravity has received increasing attention in several contexts.

Particular examples where the duality takes an important role, especially in connection with
 non perturbative properties, is the AdS/CFT correspondence \cite{ma}, related to the horizon geometry
 of p-branes and their world-volume conformal field theory description.

Another example is the connection between M-theory compactified on a
torus  T$^d$ \cite{bfss, dvv, egkr, bs, op}  and $(d+1)$
 Yang-Mills theory compactified on the dual torus $\tilde{\mbox{T}}^d$.

More closely related to the latter is the recent investigation of D-brane Born-Infeld
 actions and the role played by duality in explaining several properties of their
 Hamiltonian formulation and the corresponding energy spectrum of BPS
states \cite{hvz}. In this
 framework it is believed that Born-Infeld non abelian gauge theories with non trivial R-R
 backgrounds, are naturally described by some generalization of gauge theories on non
 commutative tori \cite{cds}.

The framework of non commutative geometry offers for instance,
   a new interpretation of the
 T-duality group $O(d,d;\Z)$ of quantum mechanical systems obtained by compactifying the
 Born-Infeld action of D-branes on T$^d$. The latter occurs in type II string theory
compactified on
 T$^d$ \cite{mz, bm, ks, do}.

These quantum mechanical systems have also been shown \cite{hb}, at least for $d\leq 4$,
 to exhibit the
 full extended U-duality symmetry\footnote{In this paper U-duality will mean both
 the classical and quantum U-duality} E$_{d+1(d+1)}$, rather than the
smaller symmetry E$_{d(d)}$ present in matrix gauge theory on T$^d$,
where it appears as an extension of the geometrical symmetry SL($d$) \cite{op, egkr}.

In previous investigations, the central charge matrix $Z$  for 0-branes played a role, not
 only as central extension of supersymmetry algebra in theories with non trivial 0-brane
 background metric, but also as effective potential of the geodesic
action of a one-dimensional Lagrangian system derived from the bulk
Einstein-Maxwell Lagrangian, in presence of moduli fields $\{\phi\}$ and
quantized charges $q_A$ of zero-branes \cite{gkk, fgk}.

The critical points of this potential were seen to determine the Bekenstein-Hawking
entropy formula as the extremization of the Weinhold potential \cite{fgk}.
\begin{equation}
W={1\over 2}\mbox{Tr}(ZZ^\dagger)
\end{equation}
or equivalently of the BPS mass $m_{BPS}=|Z_h|$ where $|Z_h|$ is the
highest eigenvalue of $\sqrt{ZZ^\dagger}$ \cite{fsf}.

In the world-volume description of 0-branes, the very same function $W$
appears as Hamiltonian of the 0-brane quantum mechanics \cite{hvz, hb, hv}
\begin{equation}
H_{\phi}(\hat q)=\sqrt{{1\over N}\mbox{Tr}(ZZ^\dagger(\phi,\hat q))}, \label{hamiltonian}
\end{equation}
where the quantized charges are replaced by a set of Hamiltonian variables $\hat q$,
 which belong to the same duality
multiplet as the quantized charges of the bulk supergravity theory
in presence of zero-brane sources.

The appearence of the central charge in the 0-brane action in
arbitrary $D=4$ supergravity backgrounds has recently been shown to
occur as a consequence of $\kappa$-supersymmetry \cite{bcfd}.

In this framework the energy spectrum of the Hamiltonian
(\ref{hamiltonian}) is given by the BPS mass formula of the effective
supergravity theory \cite{hvz}
\begin{equation}
m_{BPS}=|Z_h(\phi, q_0)|,
\end{equation}
where the hamiltonian variables $\hat q$ are replaced by their zero mode part
$q_0$ which eventually coincide with the same duality multiplet of the
quantized charges of the bulk theory, but now with the interpretation of
"fluxes" and "momenta"  of the world-volume hamiltonian description \cite{op, egkr}.

These zero modes fill representations of the U-duality group $E_{d+1(d+1)}(\Z)$
for systems with maximal supersymmetry and the BPS spectrum preserves
some fraction of supersymmetry depending on the particular orbit of the
charge vector state \cite{fm, fg, lsp}.

Note that the BPS energy $|Z_h(\phi,q_0)|$ is not the same as replacing in
$\sqrt{ZZ^\dagger(\phi, \hat q)}$ the zero mode $q_0$  of $q$, unless the states
are 1/2 BPS \cite{hvz}, which, as we will see, can only occur if the charge duality
multiplet satisfies some duality invariant conditions. At the classical
level, where the charges are continuous , this is equivalent to say that
the zero-mode part belongs to a particular orbit of the
charge vector representation of the duality group $G$.

It is the aim of the present investigation to derive general formulas
of  the energy spectrum for any torus T$^d, \; d=1,\dots 6$ and
provide a new derivation of the different BPS conditions in terms of the
U-duality invariant constraints, retrieving then the analysis of
Maldacena and one of the authors \cite{fm} as well as the classification of
Gunaydin and one of the authors \cite{fg} and Lu, Pope and Stelle \cite{lsp}.
We deal also with the case of 0-branes in theories in any dimension with 16
supersymmetries. This  is  interesting because it is related to
heterotic strings compactified on T$^d$ or Type II theories compactified on more general
manifolds (such as K$_3$).

The paper is organized as follows:

In Section 3 we consider systems compactified on T$^d, \; d=1,\dots 4$
where only 1/2 and 1/4 BPS states occur.

In Section 4 we consider the richer structure occurring for $d$=5,6 where
a complete understanding of the world volume theory is still missing.

In Section 5 the BPS conditions are derived for the case of theories
with sixteen supersymmetries in any dimension.

\section{Central charges and geometrical tools of coset spaces.}

In the present section we review the central charges for 0-branes in
theories with maximal supersymmetry and the BPS conditions on the
U-multiplet of quantized charges which entail different orbits of the
 duality group which preserve different fractions of supersymmetry.
 The analysis for theories with 16 supersymmetries
will be considered in the last section.

The  supergravity theories describing these systems can be
obtained in three different ways, by compactifying M theory on T$^{d+1}$
($(d+1)$-dimensional torus)
or type IIA and type IIB string theories on T$^{d}$. We will consider here
 supergravity theories compactified down up to $D=4$
 space-time dimensions ($d=1,\dots 6$).

Some of the results presented here overlap with previous analysis for
$d=1,\dots 4$, when only 1/2 or 1/4 BPS states are present \cite{dvv, hvz, hv}.
The analysis
of $d=5,6$ is essentially novel although the BPS conditions for 1/2, 1/4
and 1/8 BPS states were previously discussed in the literature and the
orbit classifications derived \cite{fm, fg, lsp}.

\subsection{R-Symmetry and U-duality.}

The supersymmetry algebra of type II string theory compactified on T$^d$
down to $10-d$ dimensions has an R-symmetry group and a continuous
duality group which depends on $d$. The R-symmetry is given below \cite{cr}:

\begin{center}
{\bf R-symmetry group $H$}
\begin{eqnarray}
&d=1  & \mbox{U(1)}\nonumber\\
&d=2  & \mbox{SU(2)}\times \mbox{U(1)}\nonumber\\
&d=3  & \mbox{USp(4)}\approx\mbox{O(5)}\nonumber\\
&d=4  & \mbox{USp(4)}\times\mbox{USp(4)}\approx\mbox{O(5)}\times\mbox{O(5)}\nonumber\\
&d=5  & \mbox{USp(8)}\nonumber\\
&d=6  & \mbox{SU(8)}
\end{eqnarray}
\end{center}
The U-duality groups $G$ are E$_{d+1(d+1)}$ \cite{cr}, and the R-symmetry groups are
their maximal compact subgroups. The quantum U-duality groups are  E$_{d+1(d+1)}(\Z)$
\cite{ht}.
Because of the connection between M-theory and string theory, the groups
E$_{d+1(d+1)}$ contain, both
\begin{equation}
\mbox{Gl}(d+1)\subset  \mbox{E}_{d+1(d+1)}
\end{equation}
which is the classical isometry group of the moduli space of a T$^{d+1}$
torus in M-theory and
\begin{equation}
\mbox{O}(1,1)\times\mbox{O}(d,d)\subset \mbox{E}_{d+1(d+1)}\quad (d\neq 6),\quad
\mbox{Sl}(2)\times\mbox{O}(6,6)\subset \mbox{E}_{7(7)} \quad \mbox{for}\quad d=6,
\end{equation}
which is the S-T duality group of string theory \cite{wi, adf, ht}.

In string theory the $\mbox{O}(d,d)$ group combines the geometric isometry
of the T$^d$ torus GL$(d)$ with the shift of the antisymmetric tensor
$B_{ij}\mapsto B_{ij}+N_{ij}$ while the O(1,1) factor corresponds to the
dilaton shift. The group E$_{d+1(d+1)}$ emerges from the combination of
Sl($d+1$) with O$(d,d)$ and this operation gives rise to non
perturbative symmetries which combine the N-S-NS  and R-R  fields in
supermultiplets.

The spinorial charges of the supersymmetry  algebra transform  in
representations of the R-symmetry group and
this implies that the central charges of interest to us have
 certain symmetry and reality properties.

In the case of Lorentz scalar central charges, appropriate to 0-branes,
the classification goes as follows: the central charge matrix
$Z(\phi,q)$ is in the same representation  of the R-symmetry as  the
vector fields $A_\mu$ of the corresponding theory. This gives the
following result,
\vfill\eject
\begin{center}
{\bf Central charge representation of the R-symmetry}
\begin{eqnarray}
&d=1  &\quad \mbox{{\bf 3} of O(2), ( real symmetric tensor).}\nonumber\\
&d=2  &\quad \mbox{{\bf 3(+)}  of SU(2)$\times$U(1) , (complex triplet).}\nonumber\\
&d=3  &\quad \mbox{{\bf 10} of USp(4), (real antisymmetric tensor).}\nonumber\\
&d=4  &\quad \mbox{{\bf 16} of USp(4)$\times$USp(4), (bispinor (4,4)
of O(5)$\times $O(5)).}\nonumber\\
&d=5  &\quad \mbox{{\bf 27} of USp(8), ($\Omega$-traceless symplectic antisymmetric tensor).}\nonumber\\
&d=6  &\quad \mbox{{\bf 28}  of SU(8), (complex antisymmetric tensor).}
\label{repre}
\end{eqnarray}
\end{center}

The previous results follow both, from a dynamical reduction of the 11
or 10 dimensional supergravities with 32 supercharges or by an analysis
of extended superalgebras in the appropriate dimensions \cite{to, adf2}.

Since in the original IIA theory there is only one D 0-brane (one scalar
central charge) \cite{to}, all the charges in lower dimensions come by wrapping
branes on the torus cycles, other than momenta and string windings.
In the type IIB on T$^d$, 0-branes  emerge as momenta, string windings,
and D-branes wrapped on the torus cycles.

If we want to
discuss quantum mechanical systems emerging from $d+1$ Born-Infeld
Lagrangians compactified on T$^d$, we must consider IIA D-branes
compactified on even dimensional tori and IIB  D-branes
compactified on odd dimensional tori.

The world volume description of the central charges $Z$ and quantized
charges $q$ is fairly well understood for the case of T$^d$ with
$d=1,\dots 4$. The U-duality multiplets of the quantized charges $q$
correspond to fluxes, momenta, instanton number and rank of the gauge
groups in the world volume Yang-Mills theory. The moduli dependent
central charge determines the hamiltonian  of the quantum mechanical system as well
as the energy spectrum of the BPS states \cite{hvz, egkr}.

The main role played by duality is that the central charge vector extends
the representation of the R-symmetry group to a representation of the
full duality group E$_{d+1(d+1)}$ acting on the vector field strength.
The relevant extensions are as follows \cite{cr}
\begin{center}
   {\bf{ 0-brane representation of  U-duality group}}
\begin{eqnarray}
&d=1  &\quad \mbox{{\bf 2+1} of E$_2$=SL(2)$\times$ O(1,1)}\nonumber\\
&d=2  &\quad \mbox{{\bf (3,2)} of E$_3$=Sl(3)$\times$Sl(2)}\nonumber\\
&d=3  &\quad \mbox{{\bf 10} of E$_4$=Sl(5)}\nonumber\\
&d=4  & \quad\mbox{{\bf 16} of E$_5$=O(5,5)}\nonumber\\
&d=5  &\quad \mbox{{\bf 27} of E$_{6(6)}$}\nonumber\\
&d=6  &\quad \mbox{{\bf 56} of E$_{7(7)}$}
\end{eqnarray}\label{udua}
\end{center}

The moduli space of these theories is $G/H$. At the string level this
space can be modded out further by E$_{d+1}(\Z)$ \cite{ht, wi}  something similar to
the fundamental domain versus the half plane for the prototype
Sl$(2,R)/$O(2).

In Table(2.1) we present the  U-duality multiplets for 0-branes
 in the bulk and world volume description \cite{hvz}.

The gauge fields of type II supergravity theory, as well as the Yang-Mills
world-volume fluxes complete U-duality multiplets of E$_{d+1(d+1)}$ for
$d=1,\dots 4$. These multiplets are obtained by the E$_{d(d)}$ flux and
momenta multiplets of matrix gauge theory on T$^d$ and by adding an
E$_{d(d)}$ singlet, the rank of the gauge group.  For
$d=5,6$, the Yang-Mills theory side  misses some states corresponding
to a NS five brane and K-K monopoles \cite{op, egkr}.

\begin{center}
\begin{tabular}[t]{c|c|c|}
{\bf d}& {\bf Supergravity Vector Fields} & {\bf Born-Infeld Y-M fluxes}\\
\hline
& {\bf IIA} \\
\hline
2&$ Z_\mu,g_{\mu i},b_{\mu i},A_{\mu ij}$ & $\int$Tr$F_{ij}$, $\int$
Tr$P_i$ $\int$Tr$E_i$, rank \\
\hline
4& $Z_\mu,g_{\mu i},b_{\mu i}, A_{\mu ij}$ &$\int$Tr$F_{ij}F_{kl}$,
$\int$Tr$P_i$, $\int$Tr$E_i$,$\int$Tr$F_{ij}$\\

 &$ A^D_{\mu ijkl}$& rank\\
\hline
 6&$Z_\mu,g_{\mu i},b_{\mu i}, A_{\mu ij}$&$\int$Tr$F_{ij}F_{kl}F_{pq}$,
 $\int$Tr$P_i$, $\int$Tr$E_i$, $\int$Tr$F_{ij}F_{kl}$\\

 & $A^D_{\mu ijkl},Z^D_{\mu ijklpq}$& $\int$Tr$F_{ij}$, rank\\
  &$b^{NS}_{\mu ijklp}, g^D_{\mu i}$& \\
\hline
& {\bf IIB} \\
\hline
1& $g_{\mu 1}, b_{\mu 1}, b^C_{\mu 1}$&$\int$Tr$P$, $\int$Tr$E$, rank\\
\hline
3&$g_{\mu i}, b_{\mu i}, b^C_{\mu i}, A_{\mu ijk}$& $\int$Tr$P_{i}$,
$\int$Tr$E_{i}$, $\int$Tr$F_{ij}$, rank\\
\hline
5& $g_{\mu i}, b_{\mu i}, b^C_{\mu i}, A_{\mu ijk}$&
$\int$Tr$P_{i}$,$\int$Tr$E_{i}$, $\int$Tr$F_{ij}F_{kl}$, $\int$Tr$F_{ij}$\\

 &$b^D_{\mu ijklp}$ & rank\\

 &$b^{NS}_{\mu ijklp}$ & \\
 \hline
\end{tabular}
\vskip 3mm
{\bf Table 2.1}

\end{center}
\vskip 5mm

The coset spaces $G/H$ ($G\equiv U, H\equiv R$, in our case) can be described by choosing a representative
in each equivalence class. If $\phi$ denotes the coordinates of a point in $G/H$, then the
coset representative will be given by an element $L(\phi)\in G$, in the
 equivalence class correspondig to $\phi$, that is,  $L(\phi)$ is a local section
in the principal bundle $G$ over $G/H$ with structure group $H$ . Under
the action of $g\in G$  on $G/H$ we have $\phi
\mapsto \phi_g$ and the coset representative $L(\phi)$ will be mapped to
$gL(\phi)$ which is on the fiber over $\phi_g$, so necessarily $L(\phi_g)=
gL(\phi)h(\phi)$. Taking a representation of the group $G$, (which
is also a representation of $H$) we obtain $L(\phi)$ as a matrix
$L_a^\Lambda(\phi)$ where the indexes $a$ and $\Lambda$ run in principle
over the
same representation space, the different names being used to remind the
 covariant properties of $L$. If the representation of $G$ is reducible
under $H$, one can project down  a the subspace where
the representation of $H$ is irreducible, so the index $a$ will be
understood as running
on that subspace. We will use the representations  appropriated
to 0-brane multiplets.

The central charge is  given by
\begin{equation}
Z_a(\phi,q)= (q^TL)_a=(q^T)_\Lambda L_a^\Lambda(\phi) \label{cencha}
\end{equation}
$q$ is a vector transforming under the contravariant representation of $G$,
\be
q^g=(g^{-1})^Tq
\ee
so the central charge is U-duality covariant in the sense that under a transformation
\begin{equation}
\phi\mapsto \phi_g, \quad ({q^g}^T)_\Lambda= (q^T)_\Sigma (g^{-1})^\Sigma_\Lambda,
\label{trans}
\end{equation}
then $Z\mapsto Zh$.
It then follows that any $H$-invariant function $I(Z)$
 is also U-duality invariant in the sense that
\begin{equation}
I(\phi_g,q^g)=I(\phi,q), \quad\mbox{or}\quad I(Zh)=I(Z) \label{dinva}
\end{equation}

Among the duality invariant combinations there are some which are
``topological'',i. e., they do not depend on the moduli \cite{adf3, adf4}.
This happens when
the $H$-invariant is also $G$-invariant with respect to the right action of $G$. In fact,
 since $Z=q^TL$, with $L\in G$, it is obvious that if $I(Z)$ is $G$-invariant
for $Z\mapsto Zg$, then
\begin{equation}
I(Zg)=I(Z)=I(q)
\end{equation}
Such objects exist for $d=5,6$ \cite{fsf, kk}, but not for $d=1,\dots 4$, with the implication
that a Bekenstein-Hawking entropy formula for 0-branes exist only in 4 and 5
dimensions \cite{cdcc}.

We can make a generalization of this analysis to obtain other  moduli invariant conditions.
Let us consider now  a covariant expression,
\begin{equation}
E_\alpha(Z)=E_\alpha(\phi,q),
\end{equation}
where the index  $\alpha$ runs over the space of some representation $T$ of
 $G$ (and $H$).
The covariance property means that under a left $G$  transformation

\begin{equation}
E_\alpha(Zg)=E_\beta(Z)T(g)^\beta_\alpha
\end{equation}
It follows that an equation of the form $E_\alpha(Z)=E_\alpha(\phi,q)=0$ is moduli
  independent, so $E_\alpha(q)=0$.

 Now, assume that the representation $T$  admits an $H$-invariant norm
(which is positive since $H$ is compact).
 \begin{equation}
\| E_\alpha(Z)\|^2=g^{\alpha\beta} E_\alpha E_\beta.
\end{equation}
An equation like
\begin{equation}
\| E_\alpha(Z)\|^2=0 \label{norm}
\end{equation}
is in principle moduli dependent since this expression is not
$G$-invariant. But the constraint $\|E_\alpha(Z)\|=0$ implies that
$E_\alpha(Z)=0$, which is moduli independent so $E_\alpha(q)=0$.

\medskip
The central charges satisfy some differential identities that are
inherited from the coset representatives. To see this, let us consider
the algebra valued  Maurer-Cartan form in $G$, usually expressed for a
matrix group as
\be
\alpha_{MC}=g(x)^{-1}dg(x)
\ee
where $x$ denotes coordinates on the group $G$. The components of the
Maurer-Cartan form
are left invariant forms, and one can take the pullback to $G/H$ by the
local section $L(\phi)$, giving a local,  algebra valued left invariant form on $G/H$
\be
\Omega(\phi)=L^{-1}(\phi)dL(\phi).\label{mc}
\ee
Consider now the Cartan decomposition of the Lie algebra of $G$ as
$\g=\h\oplus\k$, where $\h$ is the Lie algebra of $H$, the
maximal compact subgroup of $G$, and $\k$ can be identified with the
tangent space at the identity coset. Since our coset spaces are
symmetric spaces, the following properties are satisfied,
\be
[\h,\h]\subset\h, \quad [\h,\k]\subset\k,\quad
[\k,\k]\subset\h.\label{subal}
\ee
The second equation in (\ref{subal}) means that by the adjoint, $\h$
 acts on $\k$ as  a representation $R$ of dimension equal to dim$(G/H)$. We
can write $\Omega$ according to this decomposition of the algebra,
\be
\Omega=\omega^iT_i+P^\alpha T_\alpha
\ee
where $\{T_i\}$ form a basis of $\h$ and $\{T_\alpha\}$ form a basis of $\k$.
The projection of $\Omega$  on $\h$, $\omega^iT_i$ is a $G$ invariant
connection on the  bundle with fiber $\k$ and basis $G/H$,
 associated to the principal bundle $G(G/H)$ by the representation $R$ of $H$.
We call this bundle $E(G/H)$. The connection is expressed in an open set as
\be
(\omega)^\alpha_\beta =\omega^iC_{i\beta}^\alpha
\ee

The other components of $\Omega$, $P^\alpha$, provide us with the local
expresion of a homomorphism  between $E(G/H)$ and the tangent bundle
$T(G/H)$. By means of this homomorphism we can pull back the invariant
 Riemannian connection on $G/H$ which coincides with $\omega$. Finally,
the invariant metric on $G/H$, induced by the Cartan-Killing metric on
$G$ can be locally represented as
\be
g_{\mu\nu}=\mbox{Tr}(T_\alpha T_\beta)P_\mu^\alpha P_\nu^\beta.
\ee
(The indices ($\mu, \nu$) are 1-form indices on $G/H$).

We want now to write (\ref{mc}) using a representation of $G$ (and $H$)
labeled as we explained above indistinctely by indices of the type  $\Lambda$
or $a$, then we have
\be
dL_a^\Lambda=L_b^\Lambda \omega_a^b+L_b^\Lambda P_a^b.
\ee
By defining as usual the covariant derivative with respect to $H$
\be
\nabla_HL^\Lambda_a=dL_a^\Lambda-L_b^\Lambda \omega_a^b,
\ee
we obtain
\be
\nabla_HL^\Lambda_a=L_b^\Lambda P_a^b. \label{nabla}
\ee
Suppose now that the representation of $H$ is reducible and we want to
project onto an irreducible factor. Since $\omega_a^b$ is block diagonal,
the indices of type $a$ in $\nabla_HL^\Lambda_a$ can be understood as running on the
smaller representation, while $P_a^b$ will have in general off diagonal
components, so we could denote it by  $P_a^{b'}$, $b'$ running on the
large representation space, but still specifying the covariant
properties under $H$. This happens when matter fields are  present.
 In that case the $H$ group is
a direct product $H_R\times H_M$. $H_R$ is the R-symmetry group and
$H_M$ is some matter flavour symmetry. We assume now (as it will happen
in all our examples) that the representation of $G$ decomposes under $H$
as {\bf (1, T$_M$)}  +{\bf (T$_R$,1)}, where {\bf T$_M$} is a
representation of $H_M$ and   {\bf T$_R$} is a representation of $H_R$.
Then, there is a basis where the generic index $\Lambda$ splits into
$(a,I)$, where $a$ runs over the vector space representation of {\bf T}$_R$
and $I$ runs over the representation space of {\bf T}$_M$. Then (\ref{nabla})
decomposes as
\ba
\nabla_{H_R}L^\Lambda_a&=&L_b^\Lambda P_a^b+L_I^\Lambda P_a^I\nonumber\\
\nabla_{H_M}L^\Lambda_I&=&L_a^\Lambda P_I^a+L_J^\Lambda P_J^I,
\ea
where
\ba
\nabla_{H_R}L^\Lambda_a&=&dL^\Lambda_a - L^\Lambda_b\omega^b_a\nonumber\\
\nabla_{H_M}L^\Lambda_I&=&dL^\Lambda_I - L^\Lambda_J\omega^J_I.
\ea
The central charges $Z_a=(q^TL)_a$ and matter charges $Z_I=(q^TL)_I$
satisfy consequently  the identities
\ba
\nabla_{H_R}Z_a&=&Z_b P_a^b+Z_I P_a^I\nonumber\\
\nabla_{H_M}Z_I&=&Z_a P_I^a+Z_J P^J_I.
\label{fmc}
\ea

In the forthcoming sections we will see that these properties enter in
the discussion of the BPS conditions and their duality invariant character.

\subsection{Orbit classification of BPS states.}

In order to further study properties of central charges which will be useful in
 the following section, we would like to remind the orbit classification of
0-brane BPS configurations \cite{fg, lsp}. To do so we will state some results of matrix algebra
that will be
useful for our analysis. We consider   matrices over the real,
 complex and quaternion fields, $M^r,M^c,M^q$.
For each of these matrices, the following polar decomposition holds
 \begin{eqnarray}
M^r&=&\sqrt{M^r{M^r}^T}O\nonumber\\
M^c&=&\sqrt{M^c{M^c}^\dagger}U\nonumber\\
M^q&=&\sqrt{M^q{M^q}^\dagger}U_q
\end{eqnarray}
where the matrices $\sqrt{M{M}^\dagger}$ are hermitian and  $O,U$ and $U_q$
are orthogonal, unitary and quaternionic unitary (unitary symplectic),
 respectively.

From this decomposition it then follows that if $M^r$ is symmetric, $M^c$ hermitian
 and $M^q$ symplectic hermitian,  they can be diagonalized by an appropriate
 transformation which is respectively orthogonal, unitary and unitary symplectic.
Instead, for general matrices we can bring them to a diagonal form in the
following way,
\begin{equation}
M_D=U_1MU^\dagger_2\label{diagonal}
\end{equation}
where $U_1$ and $U_2$ belong to O(n), U(n) or USp(n) in each case.

It can also be shown that any antisymmetric matrix can be brought to a
skew-diagonal form (normal form) by a transformation \cite{kz}
\begin{equation}
M_{SD}=UMU^T
\end{equation}
where as before, $U$ belongs to the appropriate group.               .

Since the central charge vector is a 2-tensor  representation of $H$,
we can always apply
 one of the above results.

From the structure of the R-symmetry group listed above and the
 representation properties from (\ref{repre}) we will see that
 it  follows that
with an R-rotation we can always diagonalize (or skew-diagonalize) the
matrix $Z$. For $d=1,\dots 4$ there will be only two eigenvalues,
three eigenvalues for  for $d=5$ and four eigenvalues for $d=6$.

The richer structure occurring for $d=5,6$ is the  why also 1/8 BPS
states occur, instead of the two possibilities of $d=1,\dots 4$.

In the following table we list the orbits of the representations
in Eq.(\ref{udua}) corresponding to the 0-brane BPS configurations \cite{fg, lsp}.

\begin{center}
\begin{tabular}[t]{c|c|c|c|}
{\bf Orbits} &{\bf 1/2 BPS} & {\bf 1/4 BPS} &{\bf 1/8 BPS}\\
\hline
$d=1$ & Sl(2) or $\R$ & Sl(2)$\times\R $&\\
\hline
$d=2$ &${\mbox{ Sl(3)}\times\mbox{ Sl(2)}/ \mbox{Gl(2)}\propto \R^3}$ &
${\mbox{Sl(3)}\times \mbox{Sl(2)}/ \mbox{Sl(2)}\propto \R^2}$&\\
\hline
$d=3$ &$ {\mbox{Sl(5)}/(\mbox{Sl(3)}\times \mbox{Sl(2)})\propto \R^6}$ &
$ {\mbox{Sl(5)}/\mbox{O(2,3)}\propto \R^4}$&\\
\hline
$d=4$& ${\mbox{O(5,5)}/\mbox{Sl(5)}\propto\R^{10}}$ &
${\mbox{O(5,5)}/\mbox{O(3,4)}\propto \R^8}$&\\
\hline
$d=5$ & ${\mbox{E}_{6(6)}/\mbox{O(5,5)}\propto\R^{16}}$ &
${\mbox{E}_{6(6)}/\mbox{O(4,5)}\propto\R^{16}}$ &$
{\mbox{E}_{6(6)}/\mbox{F}_{4(4)}}$\\
\hline
$d=6$ & ${\mbox{E}_{7(7)}/\mbox{E}_{6(6)}\propto\R^{27}}$ &
${\mbox{E}_{7(7)}/(\mbox{O(5,6)}\propto\R^{32})\times\R}$ &
${\mbox{E}_{7(7)}/\mbox{F}_{4(4)}\propto\R^{26}},$\\
&&&$ {\mbox{E}_{7(7)}/\mbox{E}_{6(2)}}$\\
 \hline
\end{tabular}
\vskip 3mm
{\bf Table 2.2}
\end{center}
\vskip 5mm
These orbits correspond, for $d=1,\dots 4$ to the possibility of having
 two coinciding eigenvalues (1/2 BPS) or not (1/4 BPS) for the central
 charge matrix. For $d=5,6$, 1/2 BPS correspond to 3 and 4 coinciding
 eigenvalues respectively, 1/4 BPS orbits correspond to 2 equal eigenvalues
 and 2 pairs of equal eigenvalues respectively and 1/8 BPS orbits
 correspond to all different eigenvalues. In $d=6$ there are there two kinds
of 1/8 BPS orbits depending whether the quartic invariant vanishes  or not
(light-like or time-like orbit)\cite{fm, fg}.

In the following section we will see that, in spite of the fact that such
statements look moduli dependent, they are actually moduli
independent and therefore U-duality invariant, as expected from physical
considerations.

\section{BPS spectrum for 0-branes in type II string theory compactified
on T$^d$, $d=1,\dots 4$}

In the present section we consider the BPS spectrum and the central
charge matrix  for 0-branes in the cases when only 1/2 and 1/4 BPS
states exist. These are the cases where the central charge has only two
eigenvalues, as it happens for $d=1,\dots 4$.

Let us consider the relevant anticommutators, containing the scalar
central charge,
\begin{eqnarray}
d=1\quad &\{Q_i,Q_j\}&=Z_{ij}\quad (i,j=1,2);\; Z_{ij}\mbox{ real symmetric}\nonumber\\
d=2\quad &\{Q_A,Q_B\}&=Z_I\sigma_{AB}^I=Z_{AB}\quad ( A,B=1,2);\nonumber\\
&&Z^I\;\mbox{complex},\quad Z_{AB}\; \mbox{symmetric},  \nonumber\\
d=3\quad &\{Q_a,Q_b\}&=Z_{IJ}(\gamma^{IJ})_{ab}=Z_{ab}\quad
(a,b=1,\dots 4);\nonumber\\
&& Z_{ab}\; \mbox{ symmetric and symplectic} \nonumber\\
d=4\quad &\{Q_a,Q_b'\}&=Z_{ab'}\quad
(a,b'=1,\dots 4);\; Z_{ab'}\; \mbox{symplectic}
\end{eqnarray}
where $\sigma^I$ are the Pauli matrices and  $\gamma^{IJ} = 1/2 [\gamma^I
, \gamma^J ]$ ($ \gamma^I$ are the O(5) gamma matrices).
We say that  $Z_{ab}$ is a symplectic (or quaternionic) matrix if:
\begin{equation}
\bar Z =- \Omega Z \Omega \label{sym}
\end{equation}
where $\Omega$ is the bilinear form invariant under
USp(4) which satisfies
\begin{equation}
\Omega = \bar \Omega = -  \Omega^T = - \Omega^{-1}
\end{equation}
The indices $a,b$ in the gamma matrices are raised and lowered with $\Omega$.

\paragraph{ Cases $d=1,2$.} In these cases the chare matrix $Z$ is
2$\times$2 and has two independent eigenvalues. The 1/2 BPS conditions
correspond to these eigenvalues equal in magnitude.  We consider separately
both cases.

\medskip

$\bullet$ For $d=1$, $Z$ is real and symmetric. We can decompose it as
\begin{equation}
Z_{ij}=Y\delta_{ij} +Z^\alpha (T_\alpha)_{ij}
\end{equation}
where $\alpha=1,2$, $T_1=\sigma_1$ and $T_2=\sigma_3$ (the Pauli
matrices).
The characteristic equation for $Z$ is
\begin{equation}
\lambda^2 - \mbox{Tr} Z \, \lambda +\mbox{det}Z =0 \label{simple}
\end{equation}
where
\begin{eqnarray}
&& \mbox{Tr} Z = 2 Y \\
&&\mbox{det}Z = Y^2 - Z^{\alpha} Z_{\alpha}
\end{eqnarray}
It then follows that  the two solutions of (\ref{simple}) satisfy
 $|\lambda_1|=|\lambda_2|$ if either
\begin{equation}
\mbox{Tr} Z=0
\end{equation}
or
\begin{equation}
\mbox{det} Z =1/4 (\mbox{Tr}Z)^2
\end{equation}
then implying
\begin{equation}
YZ^{\alpha}Z_{\alpha} =0.\label{discri}
\end{equation}

It is obvious that  condition (\ref{discri}) is only O(2) or R-invariant,
but
the unique solution  $YZ_\alpha=0$, is an SL(2)$\times$O(1,1) or U-invariant. This is
the first example of a condition of the type (\ref{norm}). So we
retrieve the result of Ref.\cite{fm} for $d=1$, \,1/2 BPS states
\begin{equation}
q=0 \quad \mbox {or}\quad q_\alpha=0\quad
\end{equation}
\medskip

$\bullet$ In the $d=2$ case the hermiticity condition is lacking,
but still the matrix $Z^I\sigma_I$ can be diagonalized with real
eigenvalues using (\ref{diagonal}), where the difference between $U_1$
and $U_2$ is simply a phase.  We  just use  a transformation
of the R-symmetry group U(2), with SU(2) acting on the $\sigma$-matrices
and U(1) acting  as a phase on $Z^I$.

In fact, note that
\begin{equation}
Z_IZ^I=(A_I+iB_I)(A^I+iB^I)=A_IA^I -B_IB^I+2iA_IB^I
\end{equation}
Therefore, with a U(1) transformation we bring  $A_IB^I$ to zero, which
means that $\vec{A}$ and $\vec{B}$ are orthogonal vectors, so by an
orthogonal transformation we can bring them to coincide with the axes,
and only two real numbers (related to the two eigenvalues) are left.

We proceed by diagonalizing the hermitian matrix $ZZ^\dagger$. The
square root of the eigenvalues will be the eigenvalues of $Z$, that we
denote by $\lambda_1, \lambda_2$. (In this way we include both cases, when
the eigenvalues are equal and when they are opposite in sign).We have :
\begin{equation}
\mbox{Tr}ZZ^\dagger=\lambda_1^2+\lambda_2^2, \quad \mbox{Tr}(ZZ^\dagger)^2
=\lambda_1^4 +\lambda_2^4,\quad \mbox{det}ZZ^\dagger=\lambda_1^2\lambda_2^2
\end{equation}
where $\lambda_i$ are the eigenvalues obtained  as in (\ref{diagonal}).
From the characteristic equation for $ZZ^\dagger$ we have
\begin{equation}
\lambda_{1,2}^2={1\over 2}[\mbox{Tr}ZZ^\dagger\pm\sqrt{2\mbox{Tr}(ZZ^\dagger)^2
-(\mbox{Tr}ZZ^\dagger)^2}] \label{eigen}
\end{equation}
Using now the properties of the $\sigma$ matrices,
\begin{equation}
ZZ^\dagger=Z^I\sigma_I {\bar Z^J}\sigma_J=Z^I{\bar Z_I}\I +i\epsilon_{IJK}
 Z^I{\bar Z^J}\sigma^K
\end{equation}
We denote $\hat Z_K=i\epsilon_{IJK}Z^I{\bar Z^J}$. Then we have
\begin{equation}
\mbox{Tr}ZZ^\dagger=2Z^I{\bar Z_I},\quad \mbox{Tr}(ZZ^\dagger)^2
=2[(Z^I{\bar Z_I})^2+\hat{Z}_I\hat{Z}_I],
\end{equation}
Hence, the discriminant in (\ref{eigen}) is given by
\begin{equation}
\mbox{Tr}(ZZ^\dagger )^2-{1\over 2}(\mbox{Tr}ZZ^\dagger)^2=2\hat
Z^I\hat Z_I.
\end{equation}
We set $Z^I_1=A^I,\; Z^I_2=B^I$. The 1/2 BPS condition
 $\hat{Z}_I\hat{Z}^I=0$, can be written as
\begin{equation}
\|\epsilon^{\alpha \beta}Z_\alpha^IZ^J_\beta
\epsilon_{KIJ}\|=0\Rightarrow \epsilon^{\alpha \beta}Z_\alpha^IZ^J_\beta
\epsilon_{KIJ}=0.\label{epsilon}
\end{equation}
where $\|\;\|$ is the O(3)$ \times $O(2) invariant norm. As before,
 the condition obtained is  actually invariant under SL(3)$\times $SL(2), so we obtain
the  moduli independent condition of Ref.\cite{fm}:
\begin{equation}
\epsilon^{\alpha \beta}q_\alpha^Iq^J_\beta \epsilon_{KIJ}=0.
\end{equation}

\paragraph{Cases  $d=3,4$.}

In these cases the matrix $Z$ is 4-dimensional, but because of its symplectic
property (\ref{sym})
there are only two independent eigenvalues (two pairs of equal
eigenvalues), $\lambda_{1,2}$.
Indeed, we find
\begin{eqnarray}
&&\mbox{Tr}ZZ^\dagger=2(\lambda_1^2+\lambda_2^2)\nonumber\\
&&\mbox{Tr}(ZZ^\dagger)^2=2(\lambda_1^4+\lambda_2^4).
\end{eqnarray}
The characteristic equation  (or better, its square root) is
\be
\lambda^2 -{1\over 2}\mbox{Tr}ZZ^\dagger\lambda +(\mbox{det}ZZ^\dagger)^{1/2}=0,
\ee
with
\be
(\mbox{det}ZZ^\dagger)^{1/2}={1\over 8}(\mbox{Tr}ZZ^\dagger)^2-{1\over 4}
\mbox{Tr}(ZZ^\dagger)^2.
\ee
The roots are
\be
\lambda_{1,2}^2={1\over 2}\left({1\over 2}\mbox{Tr}ZZ^\dagger\pm \sqrt{\mbox{Tr}(ZZ^\dagger)^2
-{1\over 4}(\mbox{Tr}ZZ^\dagger)^2}\right)
\ee

We consider now the two cases separately,
\medskip

$\bullet$ In the $d=3$ case we can switch from Sp(4) to O(5) indices by
setting:
\begin{equation}
Z_{ab}=Z_{IJ}(\gamma^{IJ})_{ab},\quad (I,J=1,\dots 5)
\end{equation}
where
$Z_{IJ}$ is real and antisymmetric and
\begin{equation}
\gamma^{IJ}={1\over 2}[\gamma^I,\gamma^J].
\end{equation}
It is clear that $Z_{IJ}$ can be skew-diagonalized with an O(5)
transformation, so $Z_{ab}$ can be diagonalized with a USp(4) transformation.
From the relation
\begin{equation}
{1\over 2}\{ \gamma^{IJ},\gamma^{KL}\}=(g^{IJ}g^{KL}- g^{JK}g^{IL}+\epsilon^{IJKLP}\gamma_P )
\end{equation}
it follows that
\begin{equation}
ZZ^\dagger=Z^2\I + Z^P\gamma_P,
\end{equation}
where
\begin{equation}
Z^2=2Z^{PQ}Z_{PQ},\quad Z^P=\epsilon^{PIJKL}Z_{IJ}Z_{KL}.
\end{equation}
From this, we have
\ba
\mbox{Tr}ZZ^\dagger&=&4Z^2\nonumber\\
\mbox{Tr}(ZZ^\dagger)^2&=&4Z^4+4Z^PZ_P.
\ea
The 1/2 BPS condition becomes:
\begin{equation}
\mbox{Tr}(ZZ^\dagger)^2 -{1\over 4}(\mbox{Tr}ZZ^\dagger)^2=4Z^PZ_P=0
\end{equation}
which  is o(5) invariant and implies
\begin{equation}
Z^P=\epsilon^{PIJKL}Z_{IJ}Z_{KL}=0 \label{epsilon3}
\end{equation}

Equation(\ref{epsilon3}) is SL(5) invariant when $Z^{IJ}$ is in the 10-dimensional
 representation of SL(5), and therefore it is moduli independent, giving
 the result of Ref. \cite{fm},
\begin{equation}
\epsilon^{PIJKL}q_{IJ}q_{KL}=0
\end{equation}

\medskip

$\bullet$ The $d=4$ case was already discussed in Ref.\cite{dvv, hvz}, but we outline it  here for
completeness. In this case, the matrix $Z_{ab'}$ is a general O(5) bispinor.
However, since its square is hermitian it decomposes as
\begin{eqnarray}
ZZ^\dagger&=&Z^2\I +{Z_{(l)}}^P\gamma_P,\quad p=1,\dots 5\nonumber\\
Z^\dagger Z&=& Z^2\I +{Z_{(r)}}^P\gamma_P
\quad \mbox{with} \quad
 {Z_{(l)}}^P{Z_{(l)}}_P=  {Z_{(r)}}^P{Z_{(r)}}_P.
\end{eqnarray}
where the subindices $l,r$ refer to the two $O(5)$ factors of the
$R-$symmetry group.
It follows that
\begin{equation}
\mbox{Tr}ZZ^\dagger=4Z^2,\quad
\mbox{Tr}(ZZ^\dagger)^2 =4Z^4+4 {Z_{(l)}}^P{Z_{(l)}}_P.
\end{equation}
The 1/2 BPS condition is then :
\begin{equation}
\mbox{Tr}(ZZ^\dagger)^2 -{1/4}(\mbox{Tr}ZZ^\dagger)^2=4{Z_{(l)}}^P{Z_{(l)}}_P=0.
\end{equation}
The equations ${Z_{(l)}}^P{Z_{(l)}}_P= {Z_{(r)}}^P{Z_{(r)}}_P=0$ imply that the O(5)
vectors $Z^P_{(r)}$, $Z^P_{(l)}$ vanish. This is an O(5,5) invariant statement. $(Z,Z^\dagger)$
form the 16 dimensional (chiral spinor) representation of O(5,5) and the O(5,5)
10 dimensional (light-like vector) $(\mbox{Tr} \gamma ZZ^\dagger,
\mbox{Tr} \gamma Z^\dagger Z)$ then vanishes when $|Z_{(l)}|=|Z_{(r)}|=0$.
We then retrieve the condition of Ref.\cite{fm} on the quantized charges in the spinor
representation of O(5,5).

\section{BPS spectrum for the $d\!=\!5,6$ dimensional cases.}

In this section we will examine the more interesting cases of $d=5,6$,
corresponding to supergravity compactified down to $D=4,5$ dimensions
respectively.

The different BPS states, preserving some fraction of supersymmetry,
are classified by the orbits of E$_{6(6)}$ and E$_{7(7)}$
respectively as given in Table 2.2

To put this analogy in perspective it is useful to parametrize the set of BPS
 charges  allowed by the duality constraints by their eigenvalues and some
 angular variables (which can be removed by an R-symmetry transformation \cite{fg}). The
 duality constraints which follow from the BPS conditions are precisely those
 constraints which do not depend on these extra angular variables and which can
be removed by an $H$ transformation in $G$. These constraints will give
different orbits corresponding to different BPS conditions on the 0-brane
 charges.

\begin{center}
{\bf Orbits of the BPS energy levels for $d=5,6$}
\bigskip
\begin{tabular}[t]{c|c|c|c|c|}
& Orbit & dim.& eigenv. & angles\\
\hline
 d=5   \\
\hline
1/2 BPS&E$_{6(6)}$/O(5,5)$\propto\R^{16}$ & 17&1&16=dim(USp(8)/O(5)$\times$O(5)) \\
\hline
1/4 BPS&E$_{6(6)}$/O(4,5)$\propto\R^{16}$ & 26&2&24=dim(USp(8)/O(4)$\times$O(4)) \\
\hline
1/8 BPS&E$_{6(6)}$/F$_{4(4)}$ & 26+1&3&24=dim(USp(8)/USp(2)$^4$) \\
\hline
 d=6   \\
 \hline
 1/2 BPS&E$_{7(7)}$/E$_{6(6)}\propto\R^{27}$ & 28&1&27=dim(SU(8)/USp(8)) \\
 \hline
1/4 BPS&E$_{7(7)}$/(O(5,6)$\propto\R^{32})\times\R$ & 45&2&43=dim(SU(8)/USp(4)$^2$) \\
\hline
1/8 BPS&E$_{7(7)}$/F$_{4(4)}\propto\R^{26}$ & 55&4&51=dim(SU(8)/USp(2)$^4$) \\
&E$_{7(7)}$/E$_{6(2)}$& 55+1&5&51=dim(SU(8)/USp(2)$^4$) \\
 \hline
\end{tabular}
\vskip 3mm
{\bf Table 4.1}

\end{center}
\vskip 5mm

The different orbits of different BPS levels will correspond to different
 solutions of the characteristic equation of the central charge matrix (or
 its square). These different solutions will be characterized by invariant
 constraints which are moduli independent in spite of the fact that the
 eigenvalues of the matrix are moduli dependent. Becuse of this, the orbits
 are simply given by invariant constraints on the ``quantized'' charges, as found
 in Ref.\cite{fm}.

\paragraph{Case $d=6$.}

We consider the E$_7$ quartic invariant \cite{cr, adf4, kk}
\begin{equation}
I=4\mbox{Tr}(Z\bar Z)^2 -(\mbox{Tr}Z\bar Z)^2+2^4({\it Pf}Z+{\it Pf}\bar Z)
 \label{quartic}
\end{equation}
where
\begin{equation}
{\it Pf}Z={1\over 2^44!}\epsilon^{ABCDRPGH}Z_{AB}Z_{CD}Z_{RP}Z_{GH}.
\end{equation}
We want to consider second derivatives of the above quartic invariant
that could give us covariant equations. The antisymmetric matrix
$Z_{AB}$ is in the 28-dimensional representation of SU(8), while we can
express symbolically $Z_{\bf 56}=(Z_{AB},\bar Z^{AB})$, in the 56-dimensional
representation of E$_7$ (${\bf 56=28 +\bar{28}}$). Taking the  the second derivative
\begin{equation}
{\partial^2 I\over \partial Z_{\bf 56}\partial Z_{\bf 56}}\left|_{Adj_{E_7}}\right.,\label{cova}
\end{equation}
will give us a quadratic polynomial which is a symmetric tensor, in the
$({\bf 56}\times {\bf 56})_S={\bf 1596}$
representation of E$_7$ which is not irreducible and decomposes as {\bf 1463
+ 133}. {\bf 133} is the Adj$_{E_7}$, so we can project on that
 space as indicated above (\ref{cova}).
Since {\bf  133} decomposes as {\bf 63+70} under SU(8), the expression
(\ref{cova})  splits into the two following SU(8) covariant
polynomials
\begin{equation}
{\partial^2 I\over \partial Z_{AB}\bar \partial Z^{CB}}\left|_{Adj_{SU(8)}}\right.\approx
( Z_{AB}\bar  Z^{CB}-{1\over 8}\delta_A^C Z_{PQ}\bar
Z^{PQ})=V_A^C.\label{vac}
\end{equation}
\begin{equation}
{\partial^2 I\over \partial Z_{[AB}\partial Z_{CD]}}-{1\over 4!}\epsilon^{ABCDPQRS}
{\partial^2 I\over \partial \bar Z^{[AB}\partial \bar Z^{CD]}}=V^+_{[ABCD]}.
\label{selfdual}
\end{equation}
The 1/2 BPS condition is the E$_7$ invariant statement $ V_A^C=0$
and $V^+_{[ABCD]}=0$. This is the constraint
 imposed in Ref.\cite{fm} on the quantized
56 electric and magnetic charges defining a 1/2 BPS configuration.
The equation $ V_A^C=0$ implies that the matrix $ZZ^\dagger$ has four
coinciding eigenvalues (that is, it is a multiple of the identity),
while  the equation $V^+_{[ABCD]}=0$ implies
that the eigenvalues of $Z$ are real.

The vanishing of (\ref{selfdual}) follows from the vanishing of
(\ref{vac}) and the differential relations (\ref{fmc}) satisfied by $Z_{AB}$
\cite{adf3}, which in this case take the form
\be
\nabla_{SU(8)}Z_{AB}={1\over 2}\epsilon_{ABCD}\bar Z^{CD}.
\ee

We now want to consider more general cases. The
characteristic equation (or better,  its square root) is given by
\begin{equation}
\sqrt{\mbox{det}(ZZ^\dagger-\lambda\I)} =\prod_{i=1}^4(\lambda-\lambda_i)=
\lambda^4+a\lambda^3+b\lambda^2+c\lambda+d=0
\end{equation}
where
\begin{eqnarray}
a&=&-(\lambda_1+\lambda_2+ \lambda_3+ \lambda_4)\nonumber\\
&=&-{1\over2}\mbox{Tr}ZZ^\dagger\nonumber\\
b&=&\lambda_1\lambda_2+ \lambda_1\lambda_3+ \lambda_1\lambda_4+
\lambda_2\lambda_3+ \lambda_2\lambda_4+ \lambda_3\lambda_4\nonumber\\
&=&{1\over 4}[{1\over 2}(\mbox{Tr}ZZ^\dagger)^2- \mbox{Tr}(ZZ^\dagger)^2]
\nonumber\\
c&=&-(\lambda_1\lambda_2\lambda_3+\lambda_1\lambda_2\lambda_4+\lambda_1
\lambda_3\lambda_4+\lambda_2\lambda_3\lambda_4)\nonumber\\
&=&-{1\over 6}\left({1\over 8}(\mbox {Tr}ZZ^\dagger)^3 + \mbox{Tr}(ZZ^\dagger)^3-
{3\over 4} \mbox{Tr}ZZ^\dagger \mbox{Tr}(ZZ^\dagger)^2\right)\nonumber\\
d&=&\lambda_1\lambda_2\lambda_3\lambda_4\nonumber\\
&=&{1\over 4}\left({1\over 96} (\mbox{Tr}ZZ^\dagger)^4 +{1\over 8}( \mbox{Tr}
(ZZ^\dagger)^2)^2+{1\over 3} \mbox{Tr}(ZZ^\dagger)^3 \mbox{Tr}ZZ^\dagger\right.
\nonumber \\
&&\left. -{1\over 2} \mbox{Tr}(ZZ^\dagger)^4-{1\over 8} (\mbox{Tr}ZZ^\dagger)^2
 \mbox{Tr}(ZZ^\dagger)^2\right)\label{abcd}
\end{eqnarray}

 In the case of two pairs of equal roots we have
\begin{equation}
\prod_{i=1}^4(\lambda-\lambda_i)=(\lambda-\lambda_1)^2(\lambda-\lambda_2)^2.
\end{equation}
 This implies the following relations among the coefficients
 \begin{eqnarray}
 c&=&{1\over 2}a(b-{1\over 4}a^2)\nonumber\\
 d&=&{1\over 4}(b-{1\over 4}a^2)^2 \label{conditions}
 \end{eqnarray}
which imply the following relations among the invariants,
\begin{eqnarray}
&&{32\over 3}\mbox{Tr}(ZZ^\dagger)^3=4\mbox{Tr}ZZ^\dagger\mbox{Tr}(ZZ^\dagger)^2 -
{1\over 3}(\mbox{Tr}ZZ^\dagger)^3\nonumber\\
&&(\mbox{det}ZZ^\dagger)^{1/2}={1\over 64}[\mbox{Tr}(ZZ^\dagger)^2-
{1\over 4}(\mbox{Tr}ZZ^\dagger)^2]^2.\label{det}
\end{eqnarray}
The eigenvalues are given by the expression
\begin{equation}
\lambda_{1,2}={1\over 8}\mbox{Tr}ZZ^\dagger\pm{1\over 2}\sqrt{{1\over 2}
\mbox{Tr}(ZZ^\dagger)^2-{1\over 16}(\mbox{Tr}ZZ^\dagger)^2}
\end{equation}
being the BPS mass, $m_{BPS}^2$ the highest eigenvalue (+ sign).

We want now to show how the 1/4 BPS condition follows from the
E$_7$ invariance. Let us consider the E$_7$ covariant constraint
\begin{equation}
 {\partial I\over \partial Z_{AB}}=0\quad(\Rightarrow
 {\partial I\over \partial\bar Z^{AB}}=0)\label{inv1}.
 \end{equation}
 where $I$ is the invariant from (\ref{quartic}). From this, the
following quartic SU(8) invariant equations follow,
\begin{eqnarray}
 {\partial I\over \partial Z_{AB}}Z_{AB}+{\partial I\over \partial
\bar Z^{AB}} \bar Z^{AB}&=&4I=0\label{first}\\
 {\partial I\over \partial Z_{AB}}Z_{AB}-{\partial I\over \partial
 \bar Z^{AB}} \bar Z^{AB}&=&0.\label{second}
\end{eqnarray}
The second equation implies that the Pfaffian of $Z$ is real, so
\begin{equation}
{\it Pf}Z={\it Pf} Z^\dagger
\end{equation}
and therefore
\begin{equation}
({\it Pf}Z)^2=(\mbox{det}ZZ^\dagger)^{1/2}.\label{pfaf}
\end{equation}
Plugging (\ref{pfaf}) into (\ref{first}) and squaring, it
 gives $(\mbox{det}ZZ^\dagger)^{1/2}$ as in (\ref{det}).

In the same way one can show that the  equation giving $\mbox{Tr}(ZZ^\dagger)^3$ as in (\ref{det}) is
the SU(8) invariant equation
\begin{equation}
 {\partial I\over \partial Z_{AB}}{\partial I\over \partial \bar Z^{AB}}=0\label{inv2}.
 \end{equation}

In the generic case the 1/8 BPS states will correspond to 4 different
eigenvalues. They are explicitely given as follows. Define the quantities
\begin{eqnarray}
u&=& b^2 +12d-3ca\nonumber\\
v&=& 2b^3+27c^2-72bd -9abc +27 da^2\nonumber\\
w&=&\left({v+\sqrt{v^2-4u^3}\over 2}\right)^{1/3}\nonumber\\
s&=&\sqrt{{a^2\over 4}-{2b\over 3}+{u\over 3w}+{w\over 3}}
\end{eqnarray}
Then,
\begin{eqnarray}
\lambda_{1,2}=-{a\over 4}+{s\over 2}\pm{1\over 2}\sqrt{{a^2\over 2}-{4b\over 3}
-{a^3-4ab +8c\over 4s}-{u\over 3w}-{w\over 3}}\nonumber\\
\lambda_{3,4}=-{a\over 4}-{s\over 2}\pm{1\over 2}\sqrt{{a^2\over 2}-{4b\over 3}
+{a^3-4ab +8c\over 4s}-{u\over 3w}-{w\over 3}}.
\end{eqnarray}
The BPS mass, $m_{BPS}^2$ is $\lambda_1$. This is the actual
detemination of the energy spectrum for 1/8 BPS states in terms of the
duality invariant quantities (\ref{abcd}).

It is amusing that analytic expressions for the roots of a polynomial
exist only up to quartic equations, as found by Galois \cite{ed}, and
this is precisely what is required by maximal supersymmetry ($N=8$ at $D=5,6$).

\paragraph{Case $d=5$.}

The   central charge $\hat Z_{AB}$ is a symplectic, $\Omega$-traceless antisymmetric  matrix;
\begin{equation}
 \bar {\hat Z}=-\Omega\hat Z\Omega,\quad \hat Z^T=-\hat Z,\quad \mbox{Tr}\hat Z\Omega=0.
\end{equation}
This implies that the matrix
\begin{equation}
Z=\hat Z \Omega
\end{equation}
is hermitian traceless. The characteristic equation for $Z$ becomes
\begin{equation}
\sqrt{\mbox{det}Z-\lambda \I}=\prod_{i=1}^4(\lambda-\lambda_i)
=\lambda^4+b\lambda^2 +c\lambda +d=0
\end{equation}
where
\begin{eqnarray}
b&=& -{1\over 4}\mbox{Tr}Z^2\nonumber\\
c&=&  -{1\over 6}\mbox{Tr}Z^3\nonumber\\
d&=&  {1\over 8}\left({1\over 4}(\mbox{Tr}Z^2)^2-\mbox{Tr}Z^4\right)
\end{eqnarray}
A 1/4 BPS state is a state for which $c=0$. This is an E$_6$ invariant
statement since $c=I_3$ is the E$_6$ cubic invariant. In this case we get
\begin{equation}
2\lambda_{1,2}^2= {1\over 4}\mbox{Tr}Z^2\pm \sqrt{ {1\over 2}\mbox{Tr}Z^4
 -{1\over 16}(\mbox{Tr}Z^2)^2}
 \end{equation}
 The discriminant is related to the modulus of the USp(8) (and E$_{6(6)}$) vector
 \begin{equation}
 V_B^A={\partial I\over \partial Z_A^B} \approx Z^C_AZ^B_C-{1\over 8}Z^C_DZ^D_C\delta_A^B
 \end{equation}
 Indeed,
 \begin{equation}
 \mbox{Tr}V^2= \mbox{Tr}Z^4-{1\over 8} (\mbox{Tr}Z^2)^2.
 \end{equation}
 The condition for 1/2 BPS is that the discriminant vanishes. Therefore,
this implies, by positivity, $V=0$, which is an E$_6$ invariant statement
\begin{equation}
{\partial I\over \partial Z_A^B }=0
\end{equation}
We therefore have retrieved the results of  Maldacena and one of the
authors \cite{fm}.

For the 1/8 BPS state the 4 roots are given by
\begin{eqnarray}
 \lambda_{1,2}={s\over 2}\pm{1\over 2}\sqrt{{-4b\over 3}-{2c\over s}-{u\over 3w}
 -{w\over 3}}\nonumber\\
 \lambda_{3,4}=-{s\over 2}\pm{1\over 2}\sqrt{{-4b\over 3}+
 {2c\over s}-{u\over 3w}
 -{w\over 3}}
 \end{eqnarray}
 where
 \begin{eqnarray}
 u&=&b^2 +12d,\quad z=2b^3+27c^2-72bd\nonumber\\
 w&=&\left({z+\sqrt{z^2-4u^3}\over 2}\right)^{1/3}, \quad s=\sqrt{{w\over 3}
+{u\over 3w}-{2b\over 3}}
\end{eqnarray}
The BPS mass is therefore given by the highest root, $\lambda_1$.

\section{BPS  conditions for theories with 16 supersymmetries}

In this last section we will extend our analysis to theories with 16 supersymmetries.
These theories are obtained in three different ways: by compactifying Heterotic string
theory on T$^d$ ($1\leq d\leq 6$),  from M theory compactified on K$_3$ ($D=7$) and from
Type IIA
theory compactified on   K$_3$ ($D=6$).

In the theories where matter vector fields exist, the duality group $G$ depends on the matter
content and on the space-time dimension $D$. Its maximal compact subgroup   is
$H_R\times H_M$ where $H_R$ is the R-symmetry and $H_M$ is the group
acting on the matter multiplets. In our case, $H_M$=O($n$), where $n$ is
the number of matter multiplets. $G$ is of the form O(10-$D,n$)$\times$O(1,1)
for $5\leq D\leq 9$ while for $D=4$ it is SL(2)$\times$O(6,$n$). The R-symmetry  groups are
O(10-$D$) for $5\leq D\leq 9$ and O(6)$\times$O(2)$\approx$SU(4)$\times$U(1) for $D$=4. The last
result can easily been understood from the geometric symmetry of
Heterotic string on T$^6$, where $G$ is enlarged by the electric-magnetic duality for
0-branes.

The $G$ and $H_R$ representations of the 0-branes are given in the following tables.

\begin{center}
{\bf Central charge representation of $H_R$.}
\begin{eqnarray}
d=1 &\quad {\bf 1}\quad  & \mbox{O}(1)=\I\nonumber\\
d=2 &\quad {\bf 1^c}\; \mbox{complex} & \mbox{U}(1)\approx \mbox{O}(2)\nonumber\\
d=3 &\quad {\bf 3}\; \mbox{real} & \mbox{SU}(2)\approx \mbox{USp}(2)\nonumber\\
d=4 &\quad {\bf 4}\; \mbox{real} &  \mbox{O}(4)\approx \mbox{USp}(2)\times\mbox{USp}(2)\nonumber\\
d=5 &\quad {\bf 1+5}\; \mbox{real}&  \mbox{O}(5)\approx \mbox{USp}(4)\nonumber\\
d=6 &\quad {\bf 6^c}\; \mbox{complex}  & \mbox{O}(6)\times\mbox{O}(2)\approx \mbox{SU}(4)\times\mbox{U}(1)
\end{eqnarray}
\end{center}

From the above table, and according our previous analysis,
it follows that the central charge matrix $Z_a$ has only one  independent eigenvalue
for $d=1,\dots 4$ and two independent eigenvalues for $d=5,6$. Therefore, for $d=1,\dots 4$
only 1/2 BPS states can occur while for $d=5,6$ both, 1/2 and 1/4 BPS states can occur.

\begin{center}
{\bf 0-brane representation of $G$}
\begin{eqnarray}
d=1,\dots 4 & {\bf d+n}\; \mbox{real vector}& \mbox{O}(d,n)\times \mbox{O}(1,1)\nonumber\\
d=5 & {\bf (1,2)+(5\!+\! n,-1)}\; \mbox{(singlet+vector)} & \mbox{O}(5,n)\times \mbox{O}(1,1)\nonumber\\
d=6 & {\bf(2, 6\!+\!n) }\;  & \mbox{Sl}(2)\times\mbox{SO}(6,n)
\end{eqnarray} \label{diss}
\end{center}

We consider now separately the two cases $d=5,6$.

\paragraph{Case $d=5$.}

This is the case which corresponds to heterotic string on T$^5$ or M theory (Type IIA, Type IIB)
on K$_3\times$T$^2$ (K$_3\times$S$^1$). In such compactifications, $n=21$, so $G$=O(5,21)$\times$
O(1,1) but our analysis is independent of  this specific number $n$.

The central charge $\hat Z$ is an antisymmetric symplectic matrix. The hermitian matrix,
 $Z=\hat Z\Omega$ decomposes as
\begin{equation}
Z=Z^a\gamma_a +Z^0\I
\end{equation}
where $\gamma_a$ are the O(5) $\gamma$-matrices and $Z^a, Z^0$ are real.
It follows that
\begin{eqnarray}
&&\mbox{Tr}Z=4 Z^0\nonumber \\
&&(\mbox{detZ)}^{1/2}={Z^0}^2-\vec{Z}^2=  {1\over 8}(\mbox{Tr}Z)^2-{1\over 4}\mbox{Tr}Z^2
\label{trdet}\end{eqnarray}
The  characteristic equation (or better, its square root) is
\begin{equation}
\lambda^2-{1\over 2}\mbox{Tr}Z\, \lambda +(\mbox{det}Z)^{1/ 2}=0
\end{equation}
implying that  $Z$ has two coinciding eigenvalues (in absolute value) either if
\begin{equation}
\mbox{Tr}Z=0 \quad \mbox{or}\quad {1\over 4}(\mbox{Tr}Z)^2=4(\mbox{det}Z)^{1/2}
\end{equation}
Using (\ref{trdet}), the above equation directly implies
\begin{equation}
Z_0Z_a=0. \label{condi}
\end{equation}
The eigenvalues are given by
\begin{equation}
\lambda_{1,2}={1\over 2}\left({1\over 2}\mbox{Tr}Z\pm \sqrt{\mbox{Tr}Z^2-{1\over 4}
(\mbox{Tr}Z)^2}\right),
\end{equation}
being the plus sign the mass squared of the BPS state.

We discuss now the covariance of (\ref{condi}).  Since $Z_0=e^{2\sigma}m$
where $e^{2\sigma}$ parametrizes O(1,1) and $m$ is the charge associated to  $Z_0$,
$Z_0=0$ implies $m=0$ which is an O(5,n) singlet,
so it is $G$-invariant.

According to table (\ref{diss}) we write the projection of the coset representative
over the {\bf (5+n, -1)}
representation as $e^{-\sigma}L_a^\Lambda$ where $\sigma$
parametrizes O(1,1) and $L_a^\Lambda$ is the coset representative of
O(5+n)/O(5)$\times$O(n).
If $Z_I,\; I=1,\dots n$ are the matter charges associated to the $n$
matter multiplets, we have that, because of (\ref{fmc})
\be
\nabla_{O(5)}Z_a={1\over 4}\mbox{Tr}(\gamma_aP_I)Z^I -Z_ad\sigma
\ee
 therefore $Z_a=0$ implies $Z_I=0$.
This is also an O(5,n) invariant statement since, it comes by differentiating the quadratic
invariant polynomial
\begin{equation}
I=\sum_{a=1}^5Z_aZ^a- \sum_{I=1}^MZ_IZ^I.
\end{equation}
Therefore, $Z_a=Z_I=0$ implies $q^\Lambda=0$ where $q^\Lambda$,
$\Lambda=1,\dots 5+n$, is a fixed charge vector of O(5,M),
as found in \cite{fm}.

\paragraph{Case $d=6$, $(D=4)$.}

We now consider theories with 16 supersymmetries in $D=4$, as heterotic string compactified
on T$^6$, TypeII on K$_3\times$T$^2$ or M theory on   K$_3\times$T$^3$. The new phenomenon
 which occurs here is the electric-magnetic duality of 0-branes which are assigned to the
 $(2,6+n)$ representation of SU(1,1)$\times$O(6,n).

The central charge is a 4 dimensional complex matrix $Z_{AB}$, antisymmetric in the
SU(4)$\approx$O(6) indices. Therefore, the matrix
$ZZ^\dagger$ has two independent eigenvalues, given by the characteristic equation
\begin{eqnarray}
&&\left(\mbox{det}(ZZ^\dagger-\lambda \I)\right)^{1/2}=0\nonumber\\
&&\lambda^2-{1\over 2}\mbox{Tr}ZZ^\dagger\, \lambda +(\mbox{det}ZZ^\dagger)^{1/2}=0
\end{eqnarray}
with solution
\begin{equation}
\lambda_{1,2}={1\over 2}\left({1\over 2}\mbox{Tr}ZZ^\dagger\pm \sqrt{\mbox{Tr}(ZZ^\dagger))^{2}-
{1\over 4}(\mbox{Tr}ZZ^\dagger))^{2}}\right).\label{solut}
\end{equation}

A generic 1/4 BPS state has $m_{BPS}^2$ equal to the eigenvalue with +
sign above. The 1/2 BPS configuration corresponds to a vanishing discriminant, i.e.
$\lambda_1=\lambda_2$. We would like to show how this condition is SU(1,1)$\times$O(6,$n$)
invariant in the sense that it is moduli independent in spite of the fact that
the discriminant is moduli dependent.

For this purpose we proceed like for the maximally supersymmetric case in $D=4$. If the
two eigenvalues of  $ZZ^\dagger$ coincide, then the  hermitian traceless matrix
\begin{equation}
V_A^C=Z_{AC}\bar Z^{BC} - {1\over 4}\delta_A^BZ_{PQ}\bar Z^{PQ} \label{traceless}
\end{equation}
vanishes. (The discriminant is just  Tr$V^2$, the invariant norm of the SU(4) vector $V_A^C$).

Consider now the  SU(1,1)$\times$O(6,n) quartic invariant ,
\be
I=I_1^2-I_2\bar I_2
\ee
where
\begin{eqnarray}
I_1&=&Z_{AB}\bar Z^{AB}-Z_I\bar Z^I\nonumber\\
I_2&=&{1\over 4}\epsilon^{ABCD}Z_{AB}Z_{CD}-\bar Z_I\bar Z^I.
\end{eqnarray}
The fact that $I$ is an invariant was derived in Ref.\cite{adf4} and can
be  easily understood from the fact that $(I_1, I_2, \bar I_2)$  is a
triplet of SU(1,1)$\approx$O(1,2), each of the entries being O(6,n) invariant.

The equation (\ref{traceless}) can be seen as the second derivative of $I$
projected onto the adjoint representation of SU(4).
\begin{equation}
V_A^C\approx {\partial^2 I\over \partial Z_{AB}\partial \bar Z^{CD}}\left|_{Adj_{SU(4)}}\right.
\end{equation}

Indeed, let us call $U$ the ${\bf (2,6+n)}$ of Sl(2) vector
constructed with $(Z_{AB},Z_I)$ and its complex conjugate. The
quantity
\begin{equation}{\partial^2 I\over \partial U\partial U}\label{sder}
\end{equation}
is in the symmetric product $\left({\bf (2,6+n)}\times {\bf (2,6+n)}\right)\mid_S$, which
decomposes under O(6,$M$) as {\bf(3, Sym)}+{\bf(1,Adj$_{O(6,n)}$)}= {\bf(3,1)}+
 {\bf(3,TrSym)}+{\bf(1,Adj$_{O(6,n)}$)},
where {\bf Sym} is the two fold symmetric representation, {\bf TrSym} is the traceless
symmetric representation. To show that
the $V_A^C=0$ is a $G$-invariant statement we use the fact that {\bf Adj$_{O(6,n)}$}
 decomposes under
O(6)$\times$O(n) as {\bf Adj$_{O(6,n)}$}$\mapsto$ {\bf (Adj$_{O(6)}$,1)}+{\bf (1,
 Adj$_{O(n)}$)} +{\bf (6,n)}.
 We will show that the vanishing of the projection onto
 {\bf Adj$_{O(6)}$ }$\approx${\bf Adj$_{SU(4)}$}  of
(\ref{sder}) implies the vanishing of the projection onto
{\bf (1, Adj$_{O(n)}$)}  and {\bf (6,n)}. In fact, differentiating $V_A^C=0$ and
using the differential identities (\ref{fmc}) one also finds
\begin{eqnarray}
Z_I\bar Z_J -\bar Z_IZ_J=0,\label{zeta1}\\
Z_{AB}Z_J-{1\over 4}\epsilon_{ABCD}\bar Z^{CD}\bar Z_J=0.
\label{zeta2}
\end{eqnarray}
The vanishing of the three equations $V_A^C=0$, (\ref{zeta1}) and
(\ref{zeta2}) implies that the projection of (\ref{sder}) on {\bf
Ad$_{O(6,n)}$} vanishes. This
is  a SU(1,1)$\times$SO(6,$n$) invariant and therefore the moduli dependence drops out.
These three equations  can be rewritten in terms of the fixed charges $(q_\Lambda,
 p_\Lambda)$,  in the {\bf (6+n)} of O(6,$n$)  times
 the fundamental representation of Sl(2)$\approx$SU(1,1)) as
\begin{equation}
T_{\Lambda\Sigma}^{(A)}=q_\Lambda p_\Sigma -p_\Lambda q_\Sigma=0.
\end{equation}
Note that in this basis the projection  of (\ref{sder}) onto the representation
{\bf (3,Sym)} is
\begin{equation}
T_{\Lambda\Sigma}^{(S)}=(q_\Lambda q_\Sigma,\; p_\Lambda p_\Sigma,\; {1\over 2 }
(q_\Lambda p_\Sigma +p_\Lambda q_\Sigma))
\end{equation}
whose trace part is the Sl(2) triplet $(q^2.p^2,q\cdot p)$. It can be written as a matrix
\begin{equation}
T^{(0)}=\pmatrix{q^2& q\cdot p\cr
q\cdot p&p^2}.
\end{equation}
The invariant $I$ can be written either as $T_{\Lambda\Sigma}^{(A)}{T^{(A)}}^{\Lambda\Sigma}$
 or as det$T^0$, and its square root is the entropy formula for 1/4 BPS 0-branes in theories with
 sixteen supersymmetries \cite{fsf, cdcc}.

\medskip

As a final remark, let us comment on the orbits of the  O(5,n)
 and O(6,n) vectors  for BPS configurations discussed above.

 For 1/2 BPS states  at $d=5$ we have $mq_\Lambda=0$, so either $m$ or
  $q_\Lambda$ vanish. In the former case, the BPS condition requires $q_\Lambda$
 to be  time-like or light-like $(q_\Lambda q^\Lambda\geq 0)$ \cite{fm} so
the orbit is either
O(5,$n$)/O(4)$\times$O($n$) or O(5,$n$)/IO(4,$n-1$).
If $m\neq 0$ then $q_\Lambda=0$, so the orbit is a point since the
little group is O(5,n) itself.

Let us consider now the $d=6$ case. The BPS condition corresponds to the
statement that the matrix $T^{(0)}$ is positive semidefinite. This implies
\be
\mbox{det}T^{(0)}=q^2p^2-(q\cdot p)^2\geq 0,\quad \mbox{Tr}T^{(0)}=q^2
+p^2\geq 0.
\ee
From this it follows that $q^2\geq 0$ and $p^2\geq 0$.

$\mbox{det}T^{(0)}=0$ corresponds to 1/2 BPS states; this happens when
$q=\lambda p$, $(\lambda\geq 0)$.

For $\mbox{det}T^{(0)}>0$, $q^2> 0$ and $p^2>0$ and the generic 1/4 BPS
configuration will depend on five parameters, since $p,q$, by an O(2)
transformation in SL(2) can be made orthogonal $(q_\Lambda
P^\Lambda=0)$. Indeed, the first vector can be put in the form
$(p_1,0,\cdots,0,p_{n+1},0,\cdots, 0)$ and the second in the form
$(q_1, q_2,0,\cdots,0,q_{n+1},q_{n+2},0,\cdots, 0)$ by an O(6)$\times$O(n)
transformation. The orthogonality condition is used to eliminate one of
the six parameters. The remaining $7+2n$ parameters are the "angles" in
O(2)$\times$O(6)$\times$O(n)/O(4)$\times$O($n$-2). The little group in
$G$ of the two time-like vectors is O(4)$\times$O($n$).

\bigskip

\begin{center} {\bf Acknowledgements.}
\end{center}
\bigskip

We would like to thank Anna  Ceresole, Alberto  Zaffaroni and especially Raymond Stora
for enlightening discussions.


\bigskip

\begin{thebibliography}{100}

\bibitem{ma} J. Maldacena,  {\it Adv. Theor. Math. Phys.} {\bf 2} (1997) 231.
\bibitem{bfss} T. Banks, W. Fischler, S. Shenker and L. Susskind, {\it Phys.
Rev.} {\bf D55} (1997) 5112.
\bibitem{bs} D. Bigatti and L. Susskind, hep-th/9712072 (and references therein).
\bibitem{op} N. A. Obers and B. Pioline hep-th/9809039 (and references therein).
\bibitem{dvv} R. Dijkgraaf, E. Verlinde and H. Verlinde
 {\it Nucl. Phys.} {\bf B 486} (1997) 89.
\bibitem{egkr} S Elitzur, A. Giveon, D. Kutasov and E. Rabinovici
 {\it Nucl. Phys.} {\bf B 509} (1998) 122, hep-th/9707217.
\bibitem{hvz} C. Hofman, E. Verlinde and G. Zwart, hep-th/9808128.
\bibitem{cds} A. Connes, M. Douglas and A. Schwarz, {\it JHEP} 9802 (1998) 003..
\bibitem{mz} B. Morariu and B. Zumino, Proc. R. Arnowitt Fest:  A symposium on
Supersymmetry and Gravitation, College Station, 1998 (to be published by World
Scientific).
\bibitem{bm} D. Brace and B. Morariu, hep-th/9810185.
\bibitem{ks} A. Konechny and A. Schwarz, hep-th/9811159 hep-th/9901077.
\bibitem{do} M. Douglas, hep-th/9901146.
\bibitem{hb} C. Hofman and E. Verlinde, {\it JEHP} 9812 (1994) 010.
\bibitem{gkk} G.W. Gibbons, R. Kallosh and B. Kol,  {\it Phys. Rev. Lett.} {\bf 77} (1996) 4992.
\bibitem{fgk} S. Ferrara, G. W. Gibbons and R. Kallosh, {\it Nucl. Phys.} {\bf B 500} (1997)
75.
\bibitem{hv} C. Hofman and E. Verlinde, {\it JHEP} 9812 (1999) 010, and hep-th/9810219.
\bibitem{fsf} S. Ferrara, R. Kallosh and A. Strominger, {\it Phys. Rev. } {\bf
D52} (1995) 5412;\\
A. Strominger. {\it Phys. Lett. B} {\bf 383} (1996) 39;\\
S. Ferrara and R. Kallosh, {\it Phys. Rev.} {\bf D54} (1996) 1514;\\
S. Ferrara and R. Kallosh, {\it Phys. Rev.} {\bf D54} (1996) 1525.
\bibitem{bcfd} M. Bill\'o, S. Cacciatori, F. Denef, P. Fr\'e, A. Van Proeyen and D. Zanon,
hep-th/9902100.
\bibitem{fm} S. Ferrara and J. Maldacena, (19) {\it Class. Quant. Grav.} {\bf 15} (1998) 749.
\bibitem{fg} S. Ferrara and M. Gunaydin {\it Int. Jour. Mod. Phys.} {\bf A13} (1998) 2075.
\bibitem{lsp} H. Lu, K. S. Stelle and C. N. Pope, {\it Class. Quant. Grav.} {\bf
15} 537.

\bibitem{cr}  E. Cremmer, in {\it Supergravity '81}, eds. S. Ferrara and J. G.
Taylor, p. 313;\\
  B. Julia in {\it Superspace and Supergravity}, eds. S. W. Hawking and M. Rocek
  (Cambridge, 1981), p. 331.
\bibitem{ht} C.M. Hull and P.K. Townsend, {\it Nucl. Phys. } {\bf B451} (1995) 525.
\bibitem{wi} E. Witten, {\it Nucl. Phys.} {\bf B443} (1995) 85.
\bibitem{adf} L. Andrianopoli, R. D'Auria and S Ferrara, P. Fr\'e, R. Minasian, M. Trigiante, {\it Nucl. Phys.}
 {\bf B493}  (1997) 249.
\bibitem{to} P.K. Townsend, PKT Proc. 19th Johns Hopkins
Workshop on Current Problems in Particle Thoery and 5th PASCOS Interdisciplinary
Symposium, Baltimore, 1995, ed. J. Bagger (World Scientific, Singapore, 1996).
\bibitem{adf2} L. Andrianopoli, R. D'Auria and S. Ferrara, {\it Int.  Mod. Phys.}
 {\bf A13}  (1998) 431.
\bibitem{adf3} L. Andrianopoli, R. D'Auria and S. Ferrara, {\it Int. Mod. Phys.}
 {\bf A12}  (1997) 3759.
\bibitem{ed} Harold M. Edwards, {\it Galois Theory.} Springer Verlag (1984).
 \bibitem{adf4} L. Andrianopoli, R. D'Auria and S. Ferrara, {\it  Phys.
Lett. } {\bf B403}  (1997) 12.
\bibitem{kk} R. Kallosh and B. Kol, {\it Phys. Rev.} {\bf D53} (1996) 5344.
\bibitem{cdcc} M. Cvetic and C. M. Hull, {\it Nucl. Phys.} {\bf B480} (1996)
296;\\
  M.J. Duff, R.R. Khuri and J.X. Lu, {\it Phys. Rep. } {\bf 259} (1995) 213;\\
  M. Cvetic and D. Youm, {\it  Phys. Rev.} {\bf D53} (1996) 584;\\
  M. Cvetic and A. A. Tseytlin, {\it  Phys. Rev.} {\bf D53} (1996) 5619.
\bibitem{kz} Hua Loo Keng {\it  Am. J. Math.  } {\bf 66} (1944) 470;\\
  B. Zumino {\it Math. Phys.} {\bf 3} (1962) 1056.

\end{thebibliography}
\end{document}